\documentstyle[11pt,newpasp,twoside,epsfig,times,mathptm]{article}
\begin{document}
\title{Simultaneous Measurements of the Fried Parameter $r_0$ and the
       Isoplanatic Angle $\theta_0$ using SCIDAR and Adaptive Optics - First
       Results}
\author{A.R. Wei\mbox{\ss}, S. Hippler, M.E. Kasper}
\affil{Max-Planck-Institut f\"ur Astronomie, 
       K\"onigstuhl 17, 69117 
       Heidelberg, Germany}
\author{N.J. Wooder, J.C. Quartel}
\affil{Blackett Lab, Imperial College, SW72BZ London, United Kingdom}
\begin{abstract}
We present first results of a simultaneous atmospheric turbulence 
measurement campaign at the Calar Alto (Spain) 1.23\,m and 3.5\,m telescopes. 
A SCIDAR instrument and the Calar Alto Adaptive Optics system ALFA were 
used at the 1.23\,m and 3.5\,m telescopes respectively to determine essential 
turbulence parameters like the Fried parameter $r_0$, the isoplanatic 
angle $\theta_0$, and, in the case of SCIDAR, the $C_n^2$ profile 
of the atmosphere. 
Both closed-loop and open-loop data were obtained with ALFA.
The desired parameters were then calculated from this. 
This paper shows a comparison of the results obtained from the 
two different systems for several times during a period of three nights.
\end{abstract}
\section{Introduction}
The quality of correction and the overall quality of the images
achievable by the use of Adaptive Optics (AO) is strongly dependent on
the momentary properties of atmospheric turbulence as well as a good
adjustment of instrumental parameters (such as spatial and temporal
sampling rates) to these properties. A measure for the quality of the
AO correction is the shape of the point spread function (PSF) of a given
instrument at a given time. Experience shows that the PSF is subject
to rapid variations that originate in the stochastic nature of
atmospheric turbulence as well as in instrumental
limitations. Veran et al. (1997) have shown, that it is possible to
retrieve the shape of an on-axis long-exposure PSF from the measurements of a
curvature AO system. This PSF can then be used for the reconstruction
and reduction of AO compensated image data. 
Our goal is to apply a similar method to
the Calar Alto AO system ALFA and --- if possible --- reconstruct the PSF
not only on-axis but also to give an approximation of off-axis PSFs
over the whole field of view. Additionally we want to optimize the
closed-loop parameter selection process by using additional information
on the momentary state of atmospheric turbulence. For this purpose we
conducted a measurement campaign at the Calar Alto 1.23\,m and 3.5\,m
telescopes using a SCIDAR instrument and ALFA, respectively, in order to
determine the amount of information that is encoded in wave front
measurements as well as to establish the benefit of simultaneous
SCIDAR measurements when using an AO system. The two instruments will be
described more closely in the next section.   
\section{The Instruments}
\subsection{SCIDAR}
\begin{figure}
  \plotone{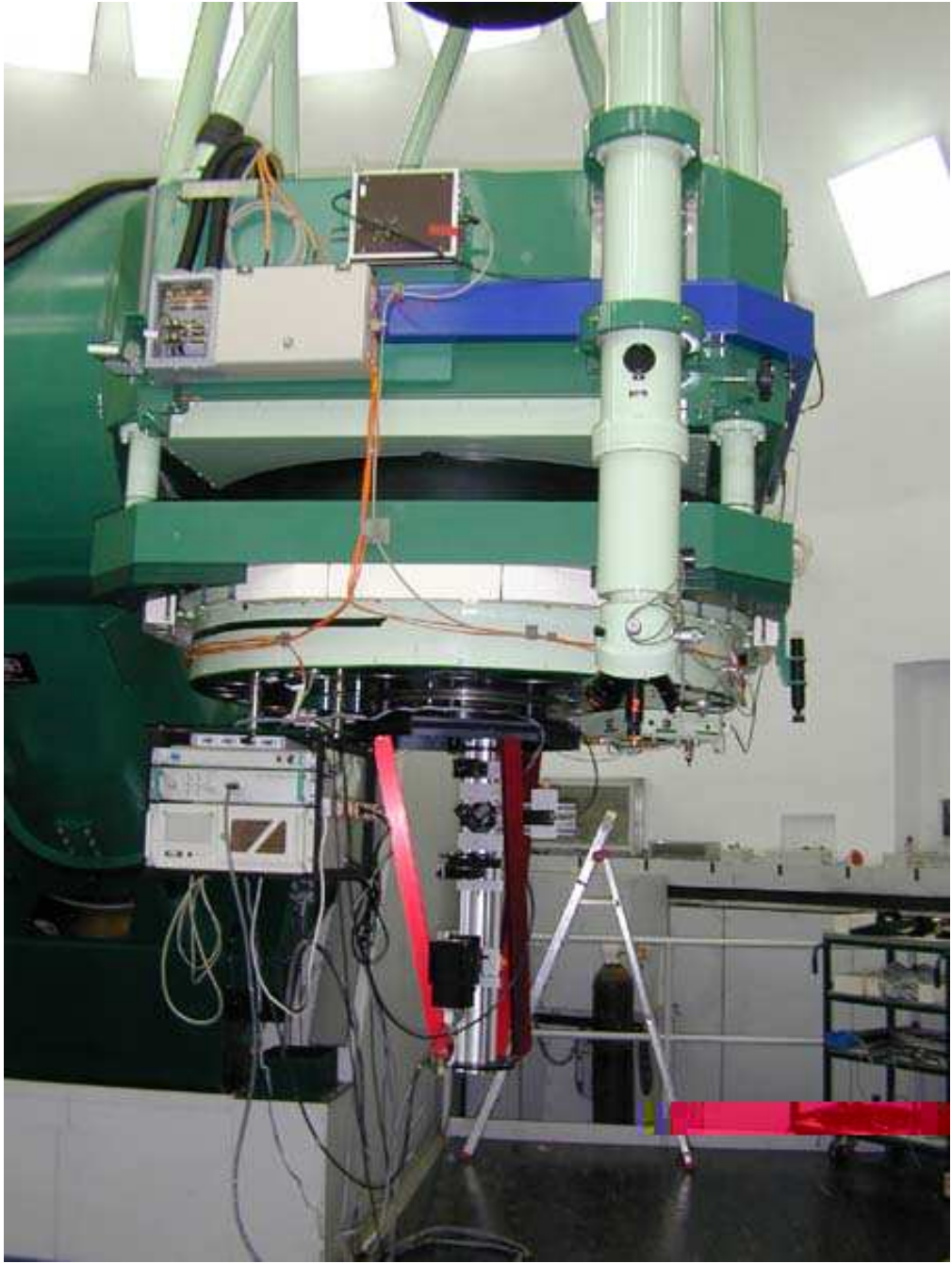}
  \caption{\label{SCIDAR123}The generalized SCIDAR instrument
  (bottom center) mounted 
  at the Cassegrain flange of the 1.23\,m telescope.}
\end{figure}

SCIDAR (Scintillation Detection And Ranging)
is a technique to determine vertical profiles of atmospheric turbulence
properties such as $C_n^2(h)$, $\tau_o(h)$ and $\vec{v}(h)$ from pupil
plane measurements of double star's scintillations (Rocca, Roddier, \&
Vernin, 1974).
While the first SCIDAR instruments measured scintillation at ground level and
were therefore not sensitive below a given height (depending on
Fresnel propagation heights) an improvement first proposed by
Fuchs, Tallon, \& Vernin (1994) that basically results in measuring scintillation
patterns virtually several kilometers {\em below} ground level makes
ground level turbulence accessible to SCIDAR measurements. This
technique is mostly referred to as {\it generalized} SCIDAR.
The generalized SCIDAR instrument used in our measurements was
developed and buiilt by Imperial College's Applied Optics 
Group at Blackett Lab in London. It has a relatively simple setup and
can therefore be easily transported. The instrument consists of a light 
intensifier that is fibre-coupled to a CCD camera. The main lens is chosen
such that pixel sampling in the pupil plane of the telescope is 
approximately $1\times1$\,cm$^2$. For the generalized SCIDAR mode, lenses can
be selected for different positions of the conjugate plane. Data are
recorded with a pixel exposure time of 1.0\,ms and a frame rate 
of 305\,Hz. Due to the high data rate and limited
disk space, the measurements are written to tape and reduced later 
(one 32 MByte block of data of each run is processed in near 
real-time in order to get an approximate $C_n^2(h)$ profile).
Figure~\ref{SCIDAR123} shows a front view of the instrument mounted on the Calar
Alto 1.23\,m telescope.
\subsection{ALFA and OMEGA-Cass}
\begin{figure}
  \plotone{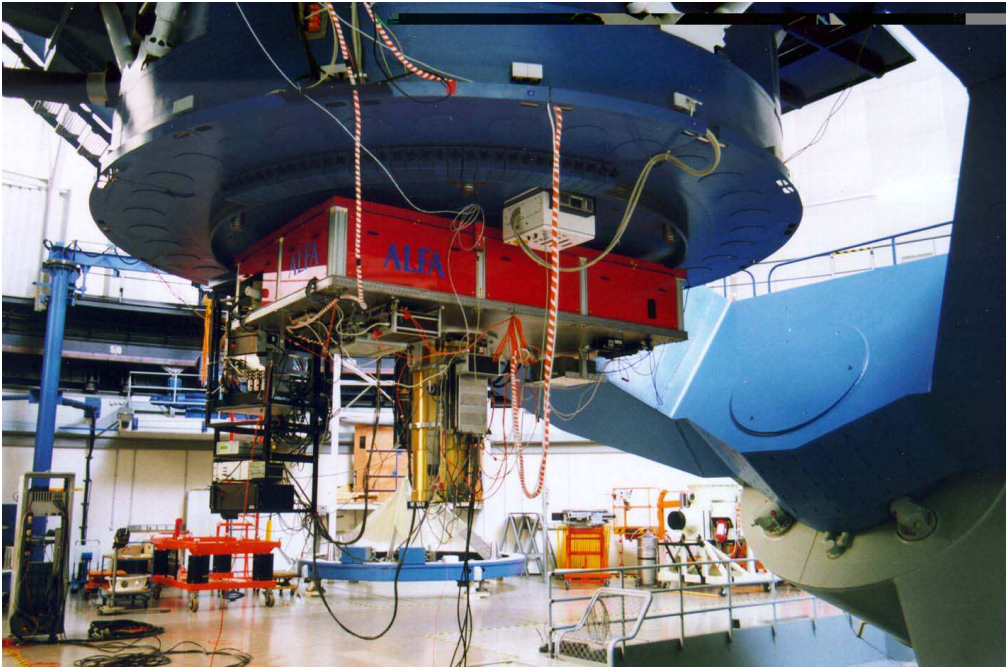}
  \caption{\label{ALFAOMEGA}ALFA (rectangular box) and the 
  near-infrared camera OMEGA-Cass (below ALFA)
  mounted at the Cassegrain flange of the 3.5\,m telescope.}
\end{figure}

ALFA is a Shack-Hartmann (SH) type AO system
(Hippler et al.1998; Kasper et al.2000). Its
AO subsystem consists of the SH Wavefront Sensor (SHWFS), a
97 actuator deformable mirror and a real-time control system based on 
digital signal processors. The SHWFS
is capable to correct wavefronts with a frequency between 25 and 1200\,Hz,
while wavefront sampling can be adjusted using a variety of lenslet
arrays (Kasper 2000). Closed-loop correction is realized by modal
control of the 97-element deformable mirror (Wirth et al. 1998) with its
conjugate plane in the telescope's pupil. Natural guide stars (NGS) or an
artificial sodium layer laser guide star (LGS) can be used as a reference
for wavefront correction. In the past we regularly achieved Strehl
ratios in the range of 20--30\% for NGSs while with LGSs a Strehl of
20\% was achieved only once so far. While ALFA in NGS mode can now be used
as a common user instrument, LGS operation remains problematic. Hence, LGSs
were not used in the measurements presented here and will in the future
only be available for test purposes.
Corrected and uncorrected images of the binaries were obtained by
MPIA's OMEGA-Cass near-infrared camera. The camera supports image plane
samplings at 0.04--0.12'' per pixel with a $1024\times1024$ near-infrared
HAWAII detector. All images here were recorded
with a sampling of 0.08'' per pixel in the infrared K-Band, hence the
diffraction limited PSF is sufficiently sampled. Figure~\ref{ALFAOMEGA}
shows ALFA and OMEGA-Cass mounted on the Calar Alto 3.5\,m telescope.
\section{The Measurements}
Measurements were done from August 31, 2000 on three consecutive
nights. While care was taken that most of the time the same binaries
were observed with both instruments, some observations were conducted simultaneously on
different objects thus enabling us to see if atmospheric parameters
are strongly dependent on the viewing direction. 
\subsection{Observed Objects}
Table~\ref{TARGETS} lists the relevant properties of the observed
binaries. The results presented here represent about 10\% of the SCIDAR 
and 30\% of the ALFA measurements taken during our run (the remaining
data sets are not yet reduced).
\begin{center}
\begin{table}
\begin{tabular}{|c|c|c|c|c|}
\hline
Object & $m_V$ & $\Delta m_V$ & Separation [``] & Sampling [km] \\
\hline\hline
95 Her &  4.3 & 0.1 & 6.3 & 0.28 \\
\hline
$\gamma$ Del & 4.1 & 1.0 & 9.6 & 0.19\\
\hline
$\gamma$ Ari & 4.8 & 0.0 & 7.8 & 0.23\\
\hline
8 Lac & 5.3 & 0.8 & 22.4 & 0.08 \\
\hline
\end{tabular}
\caption{\label{TARGETS}Properties of observed binaries and corresponding 
height sampling (for a zenith position).}
\end{table}
\end{center}
\subsection{SCIDAR measurements and Data Reduction}
SCIDAR observations were nearly continuous during the three observing 
nights. All frames used here were recorded with an approximate pixel
sampling of $1\times 1$\,cm$^2$ at a frame rate of 305\,Hz. 
The mode of observation
was usually to start observing a given binary with normal SCIDAR and to
switch lenses afterwards to get a more accurate or complete sampling of the
present turbulence structure.

\begin{figure}
  \plotone{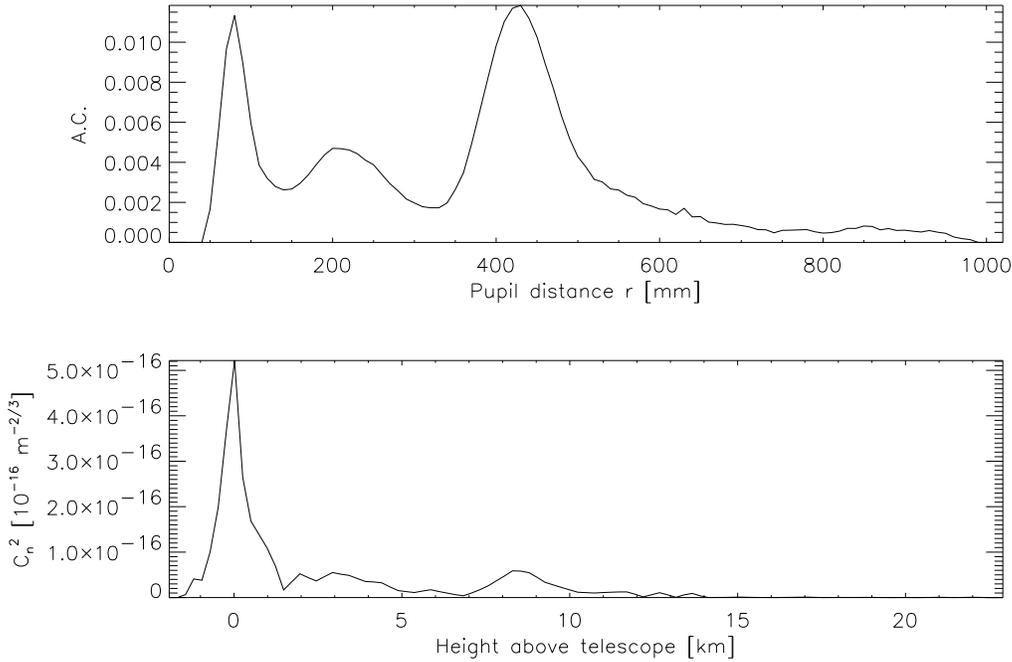}
  \caption{\label{PROFCN2}Autocorrelation profile (upper) 
                  and $C_n^2$ profile (lower) obtained from it.}
\end{figure}

Data reduction was done by a combination of the methods proposed by 
Kl\"{u}ckers et al. (1998) and Avila, Vernin, \& Cuevas (1998). 
First the autocorrelation of 
the mean-normalized frames of a run is taken and summed up. Then a section
parallel and an average section of directions not contributing to 
the line of separation of the binary's 
components are extracted and subsequently subtracted to obtain a profile
$A(r)$ that 
is both background noise reduced and has its central peak eliminated.
From this profile, the $C_n^2(h)$ structure of the atmosphere can be 
obtained by inversion of the equation:
\begin{equation}
    \label{ACHI}
A(r) = \int_{-h_0}^{\infty}C_n^2(h)K(r,h) dh
\end{equation}
where $h_0$ is the depth of the virtual conjugate plane and $K(r,h)$ is
a kernel describing the (analytical) profile produced by a layer with a
$C_n^2$ of one at altitude $h$. We used a maximum entropy method (MEM)
(Avila 2000) for
the inversion which generally showed acceptable convergence. In cases 
where MEM did not converge, a simple least-squares inversion was done.
This happened a few times for the second night due to the limited 
number of data points.
Figure~\ref{PROFCN2} 
shows an autocorrelation (A.C.) profile $A(r)$ along with the $C_n^2$ profile 
obtained from it by inversion of Eq.~\ref{ACHI}.

Given $C_n^2(h)$, the parameters we are interested in can be calculated using
(Avila et al. 1998)
\begin{equation}
  r_0 = \left[0.423\left(\frac{2\pi}{\lambda}\right)^2 
        \int dz C_n^2(z)\right]^{-3/5}
\end{equation}
and (Gonglewski et al. 1990)
\begin{equation}
    \label{TH0}
    \theta_0 = 58.1\times10^{-3}\lambda^{6/5}\left[
    \int dz C_n^2(z)z^{5/3}\right]^{-3/5}
\end{equation}
where $\lambda$ is the used wavelength and z is the height above
the telescope.

Although additional information about the turbulence, like wind speeds and a 
time constant $\tau_0$ can be extracted from our data, these parameters were
not yet calculated for this run as their influence on $r_0$ and $\theta_0$ is 
marginal. However, they do play an important role in selecting an
appropriate temporal sampling rate for AO and will thus be addressed in a 
later publication. 
\subsection{ALFA Measurements and Data Reduction}
ALFA was used to measure wavefront gradients and mode contribution both
in open-loop and closed-loop. Most of the time the same binaries as with 
SCIDAR were observed; sometimes, however, we measured a different binary
for comparison purposes. Only open-loop data was used to derive $r_0$ since
the procedure is quite straightforward. ALFA was set up to measure with
a sampling rate of 300\,Hz and a 28-subaperture (one subaperture
corresponding to a size of $\approx0.5$m on the telescope pupil) 
keystone SH lenslet array. The 
measured modes were of Karhunen-Lo\`{e}ve type (Kasper 2000). 
Typically around 20000 gradients were recorded during one measurement.
The Fried parameter $r_0$ can be determined from these measurements in two
distinct ways, one involving the gradients, the other using modal
covariances.  
Determination of $r_0$ from modal covariances involves the observation
that modal covariances scale with $(D/r_0)^{5/3}$ (Noll 1976), where $D$
is the telescope diameter; now 
an iterative scheme was applied that resulted in an estimate for $r_0$ 
(Kasper 2000).
Alternatively, $r_0$ was determined from image motion of the
subaperture spots as given by the gradients (Glindemann et al. 2000).  
\subsection{OMEGA-Cass Measurements and Data Reduction}
Along with the ALFA gradient measurements, K-Band images of the binaries 
were recorded 
on OMEGA-Cass. The pixel sampling was 0.08'' per pixel, appropriate to
the maximum obtainable K-Band resolution of 0.16'' on a 3.5\,m telescope. 
Due to the 
binaries brightness, exposure time for a single frame was 1 sec., but 
integrated frames of varying length have been used here. Only 
closed-loop images can be used to determine $\theta_0$.

In order to find an approximate value for $\theta_0$, the {\em
Strehl ratio} of the on-axis $S_{\rm on}$ and the off-axis $S_{\rm off}$ 
component of the binary is measured. Using the extended Marechal 
approximation for the Strehl ratio $S\approx\exp(-\sigma_p^2)$, where
$\sigma_p$ denotes the rms wavefront error, and the expression for the
angular isoplanatic error $<\sigma_\theta^2>=(\theta/\theta_0)^{5/3}$ 
one gets
\begin{equation}
\theta_0 = \theta \left( \sqrt{\ln{\frac{1}{S_{\rm off}}}} -
	\sqrt{\ln{\frac{1}{S_{\rm on}}}}\right)^{-6/5}
\end{equation} 
assuming linear addition of the phase rms errors 
and $\theta$ being the separation of the binary components.
\section{Results}
In the following we will shortly describe the results obtained so far. All
values have been scaled to zenith viewing direction.

\subsection{Overview}
\begin{figure}
  \plotone{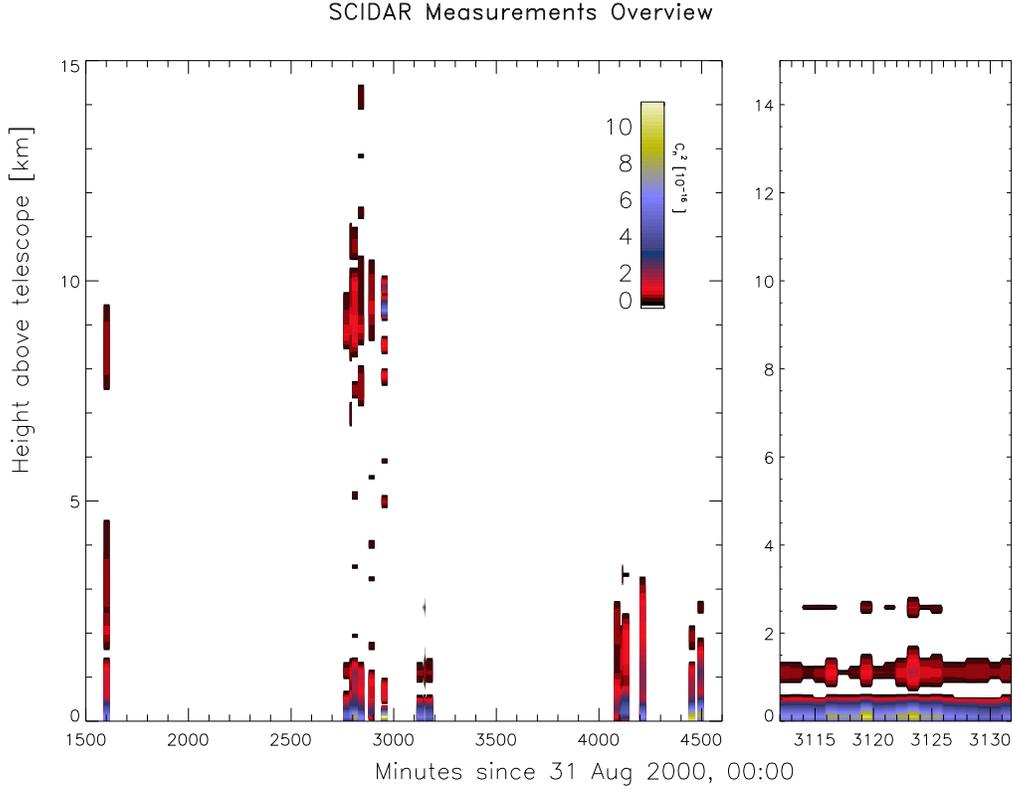}
  \caption{\label{overview}Overview of measured $C_n^2$ profiles}
\end{figure}
Figure \ref{overview} shows an overview of the measured $C_n^2$ profiles. 
The duration of each run has been exaggerated by a factor of 20 for better
readability. On the right an enlarged view of the end of the second night is
given.

Three main layers can be identified: the ground layer that is present and
dominant during all nights and which can be identified with dome seeing, if
the cross-correlations show a corresponding static component; a
second layer between 1 km and 4 km above the telescope that is fairly extended
during the first night, contracts during the second and then extends again 
in the final night (this layeer is well separated from the ground layer as 
visible in the enlarged view). Finally there is a high layer between 8 and 
11 km above the telescope that is gone in the third night. The ground layer
is the dominant layer for $r_0$; we therefore expect no significant 
differences in the behavior of the Fried parameter during the observation
period. The disappearing upper layer however should lead to an increase
of the isoplanatic angle. 
\subsection{Fried Parameter $r_0$}
\begin{figure}
  \plotone{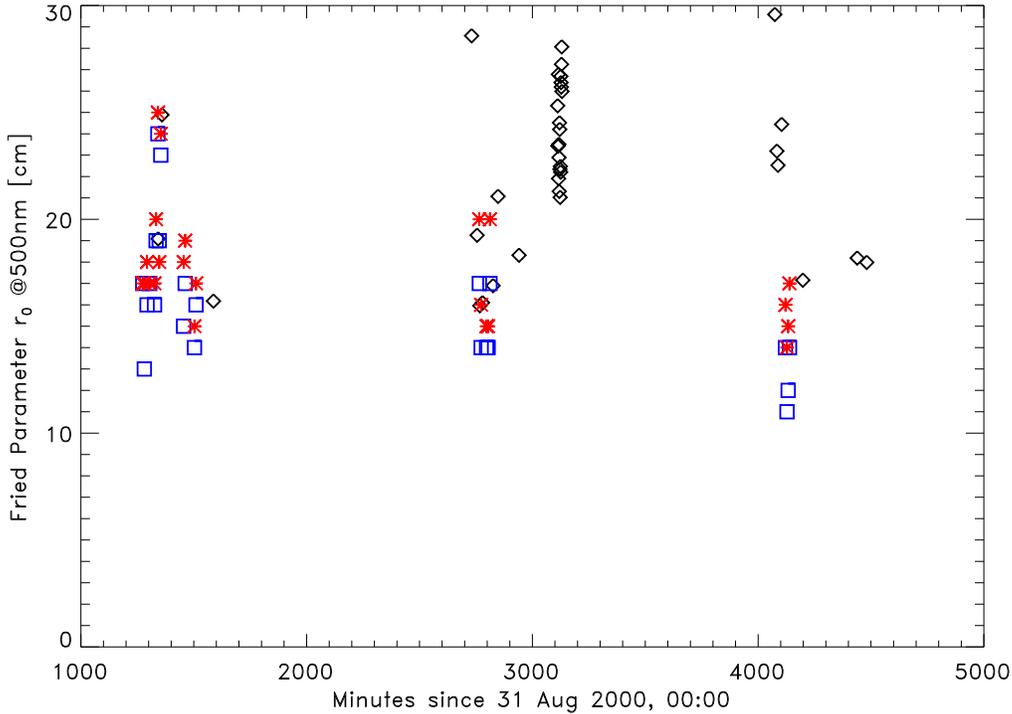}
  \caption{\label{r0}Fried parameter obtained from SCIDAR 
  measurements (diamonds), 
  ALFA modal covariances (squares) and ALFA gradients (asterisks).}
\end{figure}

Figure~\ref{r0} shows the measured values of $r_0$. A first glance shows that 
it is strongly varying during a single night which is not much of a surprise. 
Despite the quite different structure of the turbulence during the three 
nights as discussed in the previous section, the range of variation seems to
be the same. This is clearly due to the dominance of the strong ground layer 
and possible dome seeing components of $C_n^2$, the higher layers being much weaker and 
therefore only of little importance to $r_0$. A second glance reveals that 
SCIDAR $r_0$ 
values generally are a little higher than their ALFA counterparts. 
Since one would not expect the dome seeing to be worse in a larger dome this
may be a hint on a systematic overestimation of $r_0$ by generalized SCIDAR.
However, this conclusion has to wait for the reduction of all data 
sets.
\subsection{Isoplanatic Angle $\theta_0$}
The measurements of $\theta_0$ mirror the evolution of the upper turbulent 
layer, in remaining fairly stable over extended periods of time. As is to
be expected from Eq.~\ref{TH0} the presence of the upper turbulent layer 
results in a relatively low isoplanatic angle of around 22 arcseconds during 
the first night and the first part of the second night. 
With its disappearance $\theta_0$ rises to an average of about 40 arcseconds.
At a first glance the agreement between SCIDAR and OMEGA-Cass measurements seems
to be even better than for $r_0$. However, it has to be said that it turned
\begin{figure}
  \plotone{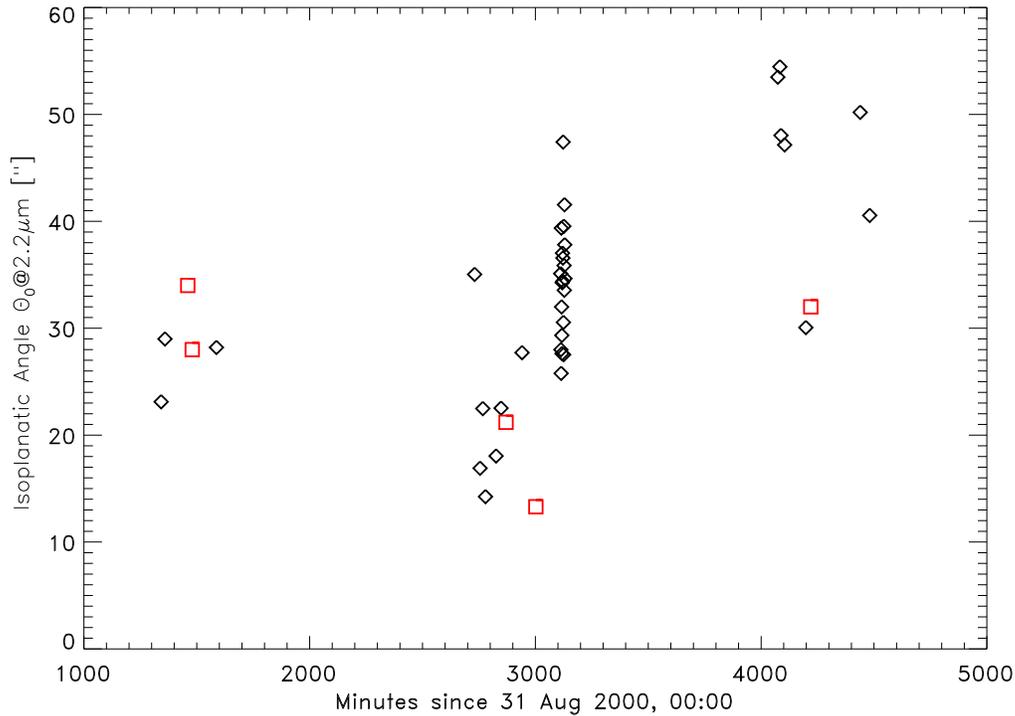}
  \caption{\label{th0}Isoplanatic angle obtained from SCIDAR measurements 
  (diamonds) and from OMEGA-Cass images (squares).}
\end{figure}
out to be extremely difficult to infer the isoplanatic angle from Strehl
ratio measurements on binaries other than 8 Lac with its large separation
of 22.4 arcseconds. The remaining images essentially yielded an infinite
$\theta_0$ which is clearly not sensible.  
\section{Conclusion and Outlook}
We have shown that SCIDAR and AO measurements of atmospheric parameters
yield similar results. Conclusions as for the use of SCIDAR in 
assisting AO observations, however, will have to wait until all
data sets are reduced. In this process a rigorous error analysis
of both ALFA and SCIDAR measurements will be done. Although there are
unfortunately only a few data points for isoplanacy measurements with
OMEGA-Cass, the tendency and magnitude seem to reproduce SCIDAR results.
This gives us hope that approximate PSFs over the whole corrected field of view
can really be found considering SCIDAR data. So the first results of
our measurement campaign look quite encouraging.
\section{Acknowledgements}
The authors thank R. Gredel and H.-W. Rix for their support in 
re-scheduling this simultaneous observation campaign on Calar Alto.

\end{document}